\documentclass[aps,prd,reprint,preprintnumbers,superscriptaddress,nofootinbib,amsmath,amssymb,floatfix]{revtex4-2}

\usepackage{graphicx, subcaption}
\usepackage{booktabs, tabularx}
\usepackage{bbold, bm, cancel, upgreek}
\usepackage{hyperref,cleveref}

\begin{document}

\preprint{UMN-TH-4021/21}

\title{Light Sterile Neutrinos and a High-Quality Axion from \\a Holographic Peccei-Quinn Mechanism}

\author{Peter Cox}
\email{peter.cox@unimelb.edu.au}
\affiliation{School of Physics, The University of Melbourne, Victoria 3010, Australia}

\author{Tony Gherghetta}
\email{tgher@umn.edu}
\affiliation{School of Physics and Astronomy, University of Minnesota, Minneapolis, Minnesota 55455 USA}

\author{Minh D. Nguyen}
\email{nguy1642@umn.edu}
\affiliation{School of Physics and Astronomy, University of Minnesota, Minneapolis, Minnesota 55455 USA}


\begin{abstract}
We present a 5D axion-neutrino model that explains the Standard Model fermion mass hierarchy and flavor structure, while simultaneously generating a high-quality axion. The axion and right-handed neutrinos transform under a 5D Peccei-Quinn gauge symmetry, and have highly suppressed profiles on the UV brane where the symmetry is explicitly broken. This setup allows neutrinos to be either Dirac, or Majorana with hierarchically small sterile neutrino masses. The axion decay constant originates from the IR scale, which in the holographically dual 4D description corresponds to the confinement scale of some new strong dynamics with a high-quality global Peccei-Quinn symmetry that produces a composite axion and light, composite sterile neutrinos. The sterile neutrinos could be observed in astrophysical or laboratory experiments, and the model predicts specific axion--neutrino couplings.
\end{abstract}

\maketitle


\section{Introduction}
\label{sec:introduction}

Two unsettled issues in the Standard Model are neutrino masses and the strong CP problem. A natural solution to the origin of the neutrino masses is the Type-I seesaw mechanism~\cite{Minkowski:1977sc,Yanagida:1979as,GellMann:1980vs,Glashow:1979nm,Mohapatra:1979ia} with Majorana masses at an intermediate scale $\gtrsim 10^{10}$ GeV. On the other hand, the most popular solution to the strong CP problem is the Peccei-Quinn (PQ) mechanism~\cite{Peccei:1977hh}, where the axion is a pseudo-Nambu-Goldstone boson~\cite{Weinberg:1977ma,Wilczek:1977pj} that results from the spontaneous breaking of a global $U(1)_{PQ}$ symmetry. These two solutions appear to be unrelated; however, the similarity of the PQ-breaking and Majorana mass scales suggests there could be an underlying mechanism responsible for both neutrino masses and the axion.

Any axion solution to the strong CP problem must address the axion quality problem, which requires extraneous, explicit violations of the global PQ symmetry to be sufficiently suppressed compared to that arising from non-perturbative QCD. Recently, a possible solution was given in Ref.~\cite{Cox:2019rro}, where the axion propagates in a slice of AdS$_5$. The PQ symmetry is gauged in the bulk, and the axion profile is suppressed near the sources of explicit PQ symmetry violation on the UV brane. The warped geometry~\cite{Randall:1999ee} can also naturally explain fermion mass hierarchies~\cite{Gherghetta:2000qt}, and a holographic DFSZ-type~\cite{Zhitnitsky:1980tq,Dine:1981rt} axion model that incorporates Standard Model flavor was presented in Ref.~\cite{Bonnefoy:2020llz}, giving predictions for flavor-violating axion--fermion couplings. The 5D framework, therefore, provides a natural setting to seek a connection between neutrino masses and the axion.

In this Letter, we extend the model of Refs.~\cite{Cox:2019rro,Bonnefoy:2020llz} to include neutrino masses. Right-handed neutrinos are introduced into the bulk and are charged under the $U(1)_{PQ}$ symmetry. The model can explain neutrinos as either Dirac or Majorana states, with hierarchies in the effective 4D neutrino Yukawa couplings and/or right-handed neutrino masses naturally generated within the 5D framework. 

The PQ symmetry forbids an explicit or spontaneously generated bulk Majorana mass for the right-handed neutrinos, leading to an accidental lepton number symmetry. However, explicit PQ (and L) violating terms are allowed on the UV brane, and the fundamental Majorana mass scale is tied to explicit Planck-scale PQ (and L) violation. Despite this connection, the sterile neutrino mass eigenstates can be naturally light. This is because the right-handed neutrino profiles can be localized towards the IR brane, away from the explicit symmetry violation. In fact, the sterile neutrino masses can range from the intermediate scale down to the eV scale (in the seesaw mechanism limit), or even lower to the theoretical Dirac limit. The model also predicts axion couplings to both the active and sterile neutrinos; however, these are well below the current experimental limits~\cite{Kelly:2018tyg}.

A connection between neutrino masses and axions was first discussed in the context of the grand unified group $SO(10)\times U(1)^\prime$, where the $U(1)_{PQ}$ symmetry is realised as a linear combination of $U(1)_{B-L}$ and $U(1)^\prime$~\cite{Mohapatra:1982tc,Holman:1982tb}. Other models based on the DFSZ axion with a connection to neutrino masses include \cite{Langacker:1986rj,Bertolini:2014aia,Clarke:2015bea}. In contrast, our 5D model automatically addresses the axion quality problem while simultaneously explaining the hierarchies of the Standard Model fermion masses and flavor structure in the quark and lepton sectors. Furthermore, by the AdS/CFT correspondence~\cite{Maldacena:1997re}, the 5D model is dual to a strongly-coupled 4D theory where the intermediate scale is dynamically generated by dimensional transmutation. The axion is identified as a composite pseudo-Nambu-Goldstone boson, and the right-handed neutrinos are also composite states. While right-handed neutrinos propagating in an extra dimension were previously considered in Refs~\cite{Grossman:1999ra,hep-ph/0110045,Huber:2000ie,hep-ph/0309218,Huber:2003sf,Gherghetta:2003hf,Agashe:2008fe,Iyer:2012db,Agashe:2015izu,1612.04810,1703.07763,Chacko:2020zze}, our setup is the first model to amalgamate neutrinos with axion physics. 


\section{Axion--Neutrino model}
\label{sec:axion-neutrino}

Consider a 5D $U(1)_{PQ}$ gauge theory with a complex scalar field $\Phi$ propagating in a slice of AdS$_5$ bounded by UV and IR branes located at $z_{UV}$ and $z_{IR}$. The metric in 5D coordinates $x^M = (x^\mu, z)$ is given by
\begin{equation}
  ds^2 = \frac{1}{(kz)^2}\left(dx^2 + dz^2\right)\equiv g_{MN} dx^M dx^N \,,
\end{equation}
where the AdS curvature scale $k \lesssim M_{P}$, with $M_{P} = 2.435 \times 10^{18}$ GeV the reduced Planck mass. The Yang-Mills--scalar action is given in \cite{Cox:2019rro}, where it is also shown that the usual global PQ symmetry that acts on the 4D axion corresponds to a particular bulk $U(1)_{PQ}$ gauge transformation.

The complex, PQ-charged scalar field $\Phi$ obtains a VEV
\begin{equation} \label{eq:eta-vev}
  \eta(z) = k^{3/2}\left(\lambda (kz)^{4-\Delta} + \sigma (kz)^\Delta\right) \,,
\end{equation}
where $\Delta$ is related to the bulk scalar mass-squared, $m_\Phi^2 = \Delta(\Delta -4) k^2$. In the dual 4D interpretation~\cite{Klebanov:1999tb} of our setup, $\sigma$ is proportional to the PQ-breaking condensate in the CFT, which has dimension $\Delta$. The coefficient $\lambda$ is associated with explicit breaking of the $U(1)_{PQ}$ symmetry on the UV brane. Note that the boundary conditions are such that the 5D gauge symmetry reduces to a global symmetry on the UV brane (guaranteeing there is no massless 4D $U(1)_{PQ}$ gauge boson), and therefore the symmetry can be explicitly broken there.\footnote{There remains an exact 5D gauge symmetry for which the gauge parameter vanishes on the UV brane.} In the limit of small 5D gauge coupling, the QCD axion is identified with the zero-mode of the phase of $\Phi$ (see Ref. [9] for details).

In Ref.~\cite{Bonnefoy:2020llz} this model was extended to include bulk fermion and Higgs fields, creating a 5D version of the DFSZ axion model~\cite{Zhitnitsky:1980tq,Dine:1981rt} that could simultaneously address both the axion quality and fermion mass hierarchy problems. Fermion zero-mode loops generate axion-gluon and axion-photon couplings localised on the UV brane, with the former leading to the usual axion solution of the strong CP problem. Here, we further extend the model to incorporate the neutrino sector. Neutrino masses are obtained by including bulk right-handed neutrinos $N_i$ $(i=1,2,3)$ with 5D Yukawa couplings and UV boundary localized Majorana masses and $\Phi$ coupling terms. The relevant part of the action is given by\footnote{Note that the delta function is defined such that $2\int_{z_{UV}}^{z_{IR}}dz\,\delta(z-z_{UV})f(z)=f(z_{UV})$.}
\begin{multline}
  S_N =-2 \int_{z_{UV}}^{z_{IR}} d^5x \sqrt{-g}\\ \times \bigg[\frac{1}{\sqrt{k}} \left( y^{(5)}_{\nu,ij} \overline{L_i} N_j H_u +\, y^{(5)}_{e,ij} \overline{L_i} E_j H_d + \mathrm{h.c.}\right) \\ 
  + \frac{1}{2}\bigg(b_{N,ij} \overline{N_i^c} N_j +\frac{y^{(5)}_{N,{ij}}}{k^{3/2}} \Phi \overline{N_i^c} N_j+ \mathrm{h.c.} \bigg)\delta(z-z_{UV})  \bigg] \,,
\label{eq:neutrinoaction}
\end{multline}
where $L_i=\big(\begin{smallmatrix} \nu_i \\ e_i \end{smallmatrix}\big)$ are the $SU(2)$ lepton doublets and $E_i$ are the $SU(2)$ singlet leptons. The 5D Yukawa couplings $y^{(5)}_{e,\nu}$ are $3\times 3$ complex matrices and $y^{(5)}_N$, $b_N$ are complex symmetric matrices. 

The model contains four $U(1)$ symmetries in the bulk: the hypercharge and PQ gauge symmetries, and accidental global lepton number and $U(1)_\Phi$ symmetries. The charges of the fields are given in \cref{tab:U(1)-charges}. The PQ charges of the Higgs fields have been chosen so that there is no mixing between the axion and the longitudinal component of the $Z$ boson, where $\tan\beta=v_u/v_d$ is the ratio of the Higgs VEVs. With these charges, the $U(1)_{PQ}$ has a boundary-localised anomaly, which is cancelled by terms on the IR boundary \cite{Bonnefoy:2020llz}.

The $U(1)_{PQ}$ and $U(1)_\Phi$ symmetries are spontaneously broken by the VEV of $\Phi$. The scalar potential on the UV boundary, given in \cite{Bonnefoy:2020llz}, includes the term $H_u H_d \Phi^2+\mathrm{h.c.}$ which explicitly breaks $U(1)_\Phi$, leaving a single pseudo-Nambu-Goldstone boson identified as the axion. As discussed previously, the $U(1)_{PQ}$ symmetry may also be explicitly broken on the UV boundary, for example by a term linear in $\Phi$. As shown in \cite{Cox:2019rro}, such terms do not disrupt the solution to the strong CP problem, provided $\Delta \gtrsim 10$. Furthermore, the lepton number and PQ symmetries are explicitly broken by $y^{(5)}_N$ and $b_N$. Note that the $U(1)_{PQ}$ gauge symmetry forbids corresponding terms in the bulk. This has important phenomenological consequences, since such terms would lead to sterile neutrino zero-mode masses of order the PQ-breaking scale, as might be expected for a high-scale seesaw. The UV boundary terms, on the other hand, can naturally give rise to hierarchically smaller sterile neutrino masses, as will be shown. These may then be accessible to experiments.

\begin{table}[t]
  \begin{tabular}{@{\hspace{0.5em}} c @{\hspace{1.5em}} c c @{\hspace{1.5em}} c @{\hspace{1.2em}} c c @{\hspace{1.5em}} c @{\hspace{0.5em}}}
    \toprule
    & $L_i$ & $E_i$ & $N_i$ & $H_u$ & $H_d$ & $\Phi$ \\
    \midrule
    $U(1)_{PQ}$ & $2\sin^2\beta$ & $4\sin^2\beta$ & 2 & $-2\cos^2\beta$ & $-2\sin^2\beta$ & 1 \\
    $U(1)_Y$ & $-\frac{1}{2}$ & $-1$ & 0 & $-\frac{1}{2}$ & $\frac{1}{2}$ & 0 \\
    $U(1)_L$ & 1 & 1 & 1 & 0 & 0 & 0 \\
    $U(1)_\Phi$ & 0 & 0 & 0 & 0 & 0 & 1 \\
    \bottomrule
  \end{tabular}
  \caption{$U(1)$ charges of the bulk fields, with $\tan\beta=v_u/v_d$.}
  \label{tab:U(1)-charges}
\end{table}

 
\subsection{Zero-mode profiles}

The equations of motion for the 5D fields can be solved via the usual expansion in Kaluza-Klein (KK) modes. The massless 4D zero-modes are then identified with the SM fermions and the axion. 

The scalar fields are parameterized as
\begin{align}
  H_u &= \frac{v_u}{\sqrt{2}}\,
  e^{\frac{i}{v_u}a_u(x^\mu,z)}
    \begin{pmatrix} 1\\0\end{pmatrix} \,, \notag \\
    H_d &= \frac{v_d}{\sqrt{2}}\,
  e^{\frac{i}{v_d} a_d(x^\mu,z)}
  \begin{pmatrix} 0\\1\end{pmatrix} \,, \notag \\
  \Phi &= \eta(z)\, e^{i a(x^\mu,z)} \,, 
  \label{eq:bulkscalarVEVs}
\end{align}
where $v_u$ ($v_d$) are the up (down)-type Higgs VEVs satisfying $(v_u^2+v_d^2)/k = v^2$, with $v\simeq 246$ GeV. The bulk and boundary scalar potentials leading to the above VEVs are given in Ref.~\cite{Bonnefoy:2020llz}. In general, $v_{u,d}$ are $z$-dependent, but for simplicity we take them to be constant, which requires a tuning between the Higgs mass terms in the bulk and on the IR boundary. Furthermore, the Higgs hierarchy problem is not addressed in the current model. The 5D fields $a_{u,d}(x,z)$ and $a(x,z)$ are the neutral Nambu-Goldstone bosons propagating in the bulk. Note that we have ignored the radial components and the electromagnetically-charged Nambu-Goldstone bosons since they play no role in the discussion. 

The equations of motion for $a_{u,d}(x,z)$ and $a(x,z)$ are coupled and the 5D fields are expanded in terms of the same set of 4D modes,
\begin{align}
  a(x^\mu,z) &= f_a^0(z) a^0(x^\mu) + \ldots \,, \notag \\
  a_{u,d}(x^\mu,z) &= f^0_{a_{u,d}}(z) a^0(x^\mu) + \ldots \,.
\end{align}
The (approximately) massless zero-mode $a^0(x^\mu)$ is identified with the axion. The profile $ f_a^0(z)$ was calculated in Ref.~\cite{Cox:2019rro}, where it was shown that explicit breaking of the PQ symmetry on the UV brane causes the profile to become suppressed by $(z/z_{IR})^\Delta$ as $z \to z_{UV}$. Away from the UV boundary the profile is approximately constant and given by
\begin{equation}
  f_a^0(z) \approx \frac{\sqrt{\Delta-1}}{\sigma_0} z_{IR} \,,
\end{equation}
where $\sigma_0=\sigma (k z_{\rm IR})^\Delta \lesssim 1$.
The profile also determines the value of the axion decay constant $F_a=f_a^0(z_{IR})^{-1}$, which is of order the IR scale, $z_{IR}^{-1}$.
 
The remaining scalar zero-mode profiles were obtained in Ref.~\cite{Bonnefoy:2020llz} and are approximately given by
\begin{multline}
  f_{a_{u,d}}^0(z) \approx v_{u,d} X_{H_{u,d}} \frac{g_5^2 k\sigma_0}{4\Delta\sqrt{\Delta-1}} \\
  \times \left[ \frac{z^2}{z_{IR}}\left( \left(\frac{z}{z_{IR}}\right)^{2(\Delta-1)} - \Delta\right) + \frac{\Delta z_{UV}^2}{z_{IR}} \right] \,,
\end{multline}
where $X_{H_{u,d}}$ are the PQ charges of the Higgs fields, and $g_5$ is the 5D $U(1)_{PQ}$ gauge coupling.

After imposing appropriate boundary conditions, the KK expansions of the 5D fermions $L_i$, ($E_i$ and $N_i$) contain left- (right-) handed 4D chiral zero-modes $L_{iL}(x^\mu)$, ($E_{iR}(x^\mu)$ and $N_{iR}(x^\mu)$) respectively. These have profiles~\cite{Gherghetta:2000qt}
\begin{align}
  f_{L_{iL}}^{0}(z) ={\cal N}_{L_i} (kz)^{2 - c_{L_i}} \,, \notag \\
  f_{E_{iR}}^{0}(z) ={\cal N}_{E_i} (kz)^{2 + c_{E_i}} \,, \notag \\
  f_{N_{iR}}^{0}(z) ={\cal N}_{N_i} (kz)^{2 + c_{N_i}} \,,
  \label{eq:fermion-profile}
\end{align}
where $c_{L_i}$, $c_{E_i}$, $c_{N_i}$ are order-one constants that parametrize the bulk lepton masses ($=c_i\, k$) and ${\cal N}_{L_i}$, ${\cal N}_{E_i}$, ${\cal N}_{N_i}$ are normalization factors. 


\subsection{Neutrino flavor structure}

The overlap of the bulk fermion profiles completely determines the neutrino flavor structure in terms of the order-one parameters $c_i$, $y_{e,\nu,N}^{(5)}$, and $b_N$. These 5D parameters can then be constrained by fitting to the two neutrino mass-squared differences and the Pontecorvo-Maki-Nakagawa-Sakata (PMNS) matrix.

Substituting \cref{eq:fermion-profile,eq:bulkscalarVEVs} into \eqref{eq:neutrinoaction} gives rise to the neutrino (zero-mode) $6\times 6$ mass matrix ${\cal M}$ defined as
\begin{equation}
  -\frac{1}{2} \int d^4x \, \begin{pmatrix}
  \overline{\vec{\nu}_L} & \overline{\vec{N}_R^c}
  \end{pmatrix}
  \begin{pmatrix} 0 & m_D \\ m_D^T & M_{M}\end{pmatrix}
  \begin{pmatrix}
  \vec{\nu}_L^c \\
  \vec{N}_R
  \end{pmatrix}
  + \mathrm{h.c.}\,,
  \label{eq:neutrinomassmatrix}
\end{equation}
with $(\vec{\nu}_L^c,\vec{N}_R)\equiv(\nu_{1L}^c,\nu_{2L}^c,\nu_{3L}^c,N_{1R},N_{2R},N_{3R})$. The Dirac mass matrix is given by
\begin{align} 
  m_D^{ij} &= y_{\nu,ij}^{(5)} \frac{\sqrt{2}v_u}{\sqrt{k}} \int_{z_{UV}}^{z_{IR}} \frac{dz}{(kz)^5}~f_{L_{iL}}^0(z) f_{N_{jR}}^0(z) \,, \label{eq:Dirac-mass} \\
  &\simeq \frac{v_u\sqrt{(-1+2c_{L_i})(1+2c_{N_j})}}{\sqrt{2k}(c_{L_i}-c_{N_j})} (kz_{IR})^{-\frac{1}{2}-\min(c_{L_i},c_{N_j})} \,,
  \label{eq:Dirac-mass-approx}
\end{align}
and, taking $kz_{UV}=1$, the Majorana mass matrix is
\begin{align} 
  M_M^{ij} &= \left(y_{N,ij}^{(5)} (\lambda+\sigma) + b_{N,ij} \right) (f_{N_{iR}}^0(z_{UV}))^2 \,, \label{eq:Majorana-mass} \\
  &\simeq k \,\hat{y}_{N,ij} \left(c_{N_i}+\frac{1}{2}\right)(k z_{IR})^{-1-2c_{N_i}} \,, 
  \label{eq:Majorana-mass-approx}
\end{align}
with $\hat{y}_{N,ij}\equiv y^{(5)}_{N,ij}(\lambda+\sigma)+b_{N,ij}$. Eqs.~\eqref{eq:Dirac-mass-approx} and \eqref{eq:Majorana-mass-approx} also assume $z_{UV}\ll z_{IR}$, $c_{L_i}>1/2$, and $c_{N_i}>-1/2$. Notice that for $c_{N_i}>0$, the effective Majorana masses are suppressed relative to the PQ-breaking or IR scale. For $c_{N_i}<0$, on the other hand, the right-handed neutrinos have masses of order the IR scale, and can no longer be treated as approximately massless modes with profiles given by \cref{eq:fermion-profile}. We therefore restrict our discussion to $c_{N_i}>0$. This corresponds to right-handed neutrinos that are localized towards the IR brane and therefore in the dual 4D description are mostly composite. For $c_{L_i}>0$, the left-handed neutrinos are mostly elementary. The Majorana mass matrix \eqref{eq:Majorana-mass} can be made diagonal and non-negative via a unitary rotation of the $N_i$, and we work in this basis where $\hat{y}_N$ is diagonal.

The neutrino mass matrix in \eqref{eq:neutrinomassmatrix} is diagonalized by a unitary matrix $\cal U$ to give ${\cal U}^T {\cal M} {\cal U} = {\rm diag}(m_{\upnu_i})$, where $m_{\upnu_i}$ are the six neutrino mass eigenvalues. The mass eigenstates $\upnu_i$ are 
\begin{equation}
  \vec{\upnu} = \mathcal{U}^\dagger \begin{pmatrix} \vec{\nu}_L^c \\ \vec{N}_R \end{pmatrix} + \mathcal{U}^T \begin{pmatrix} \vec{\nu}_L \\ \vec{N}_R^c \end{pmatrix} \,.
\end{equation}

In the seesaw limit\footnote{The Frobenius norm $\|A\|=\big[\sum\limits_{i,j}(A_{ij})^2\big]^{1/2}$, for a matrix $A$.}, $\|M_M^{-1} m_D^T\|\ll1$, the six eigenstates split into two distinct sets. One set contains the “active” light neutrinos that are mostly $SU(2)$ doublets and the other set contains the heavier ``sterile" neutrinos that are mostly Standard Model gauge singlets. The mixing matrix is then approximately~\cite{Dreiner:2008tw}
\begin{equation}
  \mathcal{U} \simeq
  \begin{pmatrix}
    \left(\mathbb{1} - \frac{1}{2} \Theta^\dagger \Theta \right) U_\nu & \Theta^\dagger U_N \\
    -\Theta U_\nu & \left(\mathbb{1} - \frac{1}{2} \Theta \Theta^\dagger\right) U_N
  \end{pmatrix} + \mathcal{O}(\|\Theta\|^3) \,,
  \label{eq:mixing-matrix}
\end{equation}
where $\Theta=M_M^{-1} m_D^T$, and $U_\nu$, $U_N$ are the $3\times3$ matrices that diagonalize the active and sterile neutrinos respectively. To leading order, the active masses are $\mathrm{diag}(m_{\nu_i}) \simeq - U_\nu^T (m_D M_M^{-1} m_D^T) U_\nu$, while the sterile masses are $\mathrm{diag}(m_{N_i}) \simeq M_M$. The PMNS matrix is $V_{PMNS} = A_L^{e\dagger} U_\nu$, where $A_L^{e\dagger}$ is the unitary matrix that rotates the left-handed charged leptons to the mass basis. After using the phase freedom of the charged leptons to remove three redundant phases, it can be expressed using the standard parameterization.

Interestingly, when $c_{L_i}>c_{N_j}\, \forall\, i,j$, the active neutrino masses do not depend on the IR scale. This is due to the correlation between the effective Dirac and Majorana masses in \cref{eq:Dirac-mass-approx,eq:Majorana-mass-approx}, such that \linebreak $m_\nu \propto v^2/k$. The 4D dual description provides an alternative viewpoint on the neutrino mass mechanism in which there are elementary, Planck-scale Majorana fermions that mix with the composite right-handed neutrinos. Including the (mostly) elementary left-handed neutrinos, the active neutrino masses then arise from a seesaw mechanism.

We also consider the possibility that neutrinos are Dirac fermions ($M_M=0$). In this case it is convenient to instead define the mass eigenstate Dirac fermions by
\begin{equation}
  \upnu =  A_L^{\upnu\dagger} \nu_L + A_R^{\upnu\dagger} N_R \,,
\end{equation}
where $A_L^{\upnu\dagger} m_D A_R^\upnu = {\rm diag}(m_{\upnu_i})$, and $V_{PMNS} = A_L^{e\dagger} A_R^\upnu$. Note that the Dirac limit does not necessarily require that lepton number is preserved on the UV brane. Instead, the Dirac limit can actually be obtained while assuming Planck-scale lepton number violation on the UV brane~\cite{Gherghetta:2003hf}. By formally taking $c_N\gg 1$ in \eqref{eq:Majorana-mass-approx}, the effective 4D Majorana mass $M_M\rightarrow 0$. For simplicity, we will not consider this pseudo-Dirac limit and instead just analyse the Majorana (seesaw-mechanism) limit and the pure Dirac limit.


\subsection{Axion--neutrino couplings}

The axion--neutrino couplings are obtained by first removing the $a_u$-dependence in \cref{eq:neutrinoaction} via a 5D field redefinition of the form
\begin{align} 
  \nu_i(x,z) \rightarrow & \,e^{i (\omega+\frac{1}{2})\frac{1}{v_u}a_u(x,z)} \nu_i(x,z) \,, \nonumber \\ 
  N_i(x,z) \rightarrow & \,e^{i(\omega-\frac{1}{2})\frac{1}{v_u} a_u(x,z)} N_i(x,z) \,,
  \label{eq:5Dfermionredef}
\end{align}
where $\omega$ is an arbitrary real parameter. The 5D fermion kinetic and boundary $N_i$ terms are not invariant under this transformation, giving rise to
\begin{multline}
  \int_{z_{UV}}^{z_{IR}} \frac{d^5x}{(kz)^4} \bigg[ i\left(\partial_M \frac{a_u}{v_u} \right)\bigg( \bar{N}_i\gamma^M N_i - \bar{\nu}_i\gamma^M \nu_i \\
  - 2\omega \left( \bar{N}_i\gamma^M N_i + \bar{\nu}_i\gamma^M \nu_i \right) \bigg) \\
  -\bigg( \Big(b_{N,ij} + \frac{y^{(5)}_{N,{ij}}}{k^{3/2}} \Phi \Big) \overline{N_i^c} N_j e^{i(2\omega-1)\frac{a_u}{v_u}} + \mathrm{h.c.} \bigg)\delta(z-z_{UV}) \bigg] \,.
  \label{eq:deltaS}
\end{multline}
The terms proportional to $\omega$ do not contribute to the S-matrix. This is seen by identifying the lepton number current in the second line of \eqref{eq:deltaS}. The (anomalous) Ward-Takahashi identity for lepton number then guarantees that contributions to the S-matrix from the $\omega$-dependent terms in the second and third lines cancel.

Restricting our focus to the zero-modes, the axion--neutrino couplings are then
\begin{multline}
  i \int_{z_{UV}}^{z_{IR}} \frac{d^5x}{(kz)^4} (\partial_\mu a^0) \frac{f_{a_u}^0(z)}{v_u}\bigg[ \bar{N}_{iR} (f^0_{N_{iR}}(z))^2 \gamma^\mu N_{iR} \\ 
  - \bar{\nu}_{iL} (f^0_{L_{iL}}(z))^2 \gamma^\mu \nu_{iL} \bigg] \,.
\end{multline}
We have neglected additional UV boundary terms, which lead to axion--sterile neutrino couplings, since they are highly suppressed by the $f_a$ or $f_{a_u}$ profile at $z=z_{UV}$. Integrating over the profiles and rotating to the fermion mass basis we obtain the 4D effective action
\begin{equation} \label{eq:effective-action}
  S_{4D} \supset i\int d^4x\, \frac{\partial_\mu a^0}{2F_a} \bar{\upnu}_i \gamma^\mu \left( (c_\upnu^V)_{ij} - (c_\upnu^A)_{ij} \gamma^5 \right) \upnu_j \,,
\end{equation}
where $\gamma^5 = {\rm diag}(\mathbb{1},-\mathbb{1})$. The vector and axial-vector couplings are given by $c_\upnu^V=i\,\mathrm{Im}(\xi_\upnu)$ and $c_\upnu^A=\mathrm{Re}(\xi_\upnu)$, where
\begin{equation} \label{eq:cnu_bulk}
  \frac{(\xi_\upnu)_{ij}}{F_a} = \int_{z_{UV}}^{z_{IR}} \frac{dz}{(kz)^4} \, \frac{f_{a_u}^0(z)}{v_u}~ \mathcal{U}^\dagger_{ik} \, \mathcal{F}^{kk} \, \mathcal{U}_{kj} \,,
\end{equation}
with $\mathcal{F}=\text{diag}((f^0_{L_{1L}})^2,(f^0_{L_{2L}})^2,(f^0_{L_{3L}})^2,(f^0_{N_{1R}})^2,(f^0_{N_{2R}})^2,\allowbreak(f^0_{N_{3R}})^2 )$. Note that the vector (axial-vector) couplings in \cref{eq:effective-action} are symmetric (anti-symmetric).

In the Dirac neutrino case, the axion--neutrino couplings are similar to those obtained for the charged fermions in \cite{Bonnefoy:2020llz}, and can be written as
\begin{multline} 
  \frac{(c_\upnu^{V,A})_{ij}}{F_a} 
  = \int_{z_{UV}}^{z_{IR}} \frac{dz}{(kz)^4}\, \frac{f_{a_u}^0(z)}{v_u} \\ \times\left( ({A^\upnu_R}^\dagger)_{ik}  (f^0_{N_{kR}})^2 (A^\upnu_R)_{kj} \mp ({A^\upnu_L}^\dagger)_{ik} (f^0_{L_{kL}})^2 (A^\upnu_L)_{kj}  \right) \,.
  \label{eq:cVA_Dirac}
\end{multline}

Finally, note that for on-shell fermions the corresponding matrix elements are proportional to $(c^{V,A})_{ij} (m_i \mp m_j)$, so that the axion--active neutrino couplings can be neglected in most cases.


\section{Phenomenology} 

\begin{figure*}[t]
  \centering
  \includegraphics[width=0.458\textwidth]{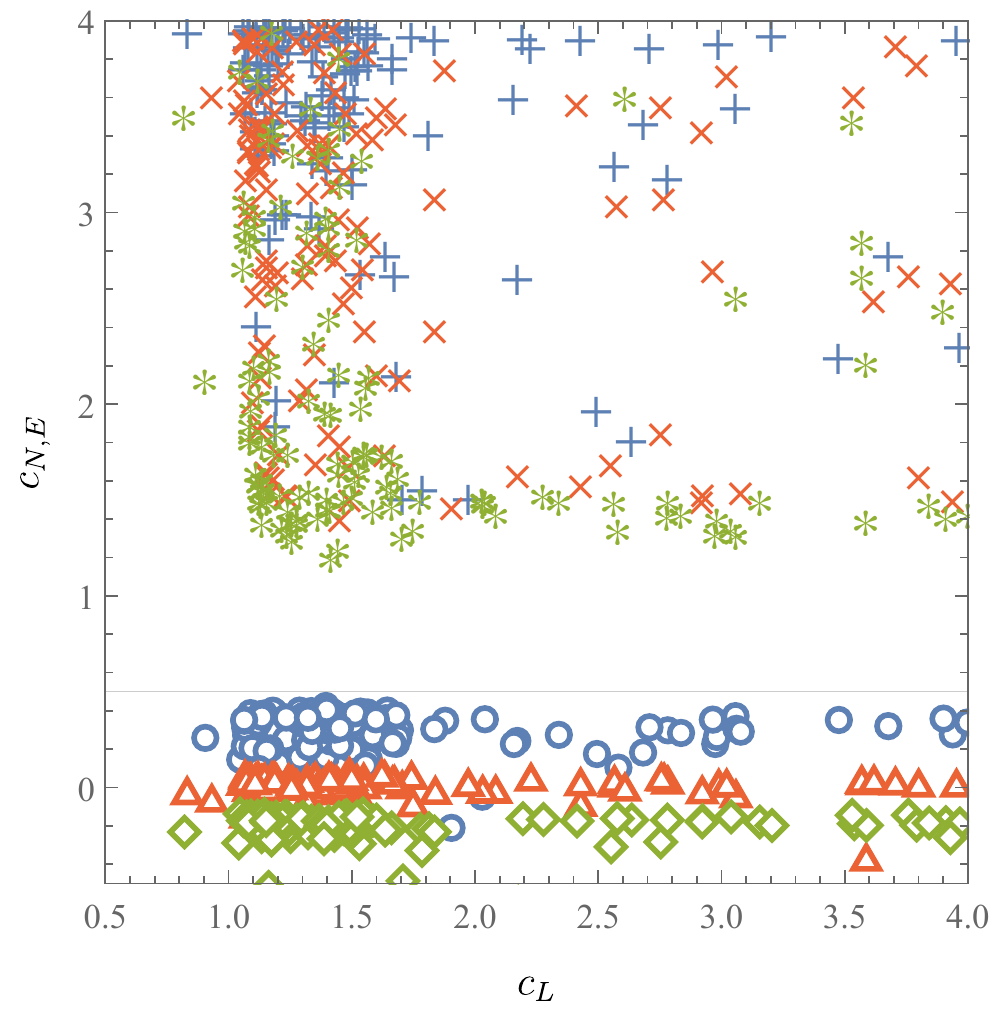}
  \includegraphics[width=0.48\textwidth]{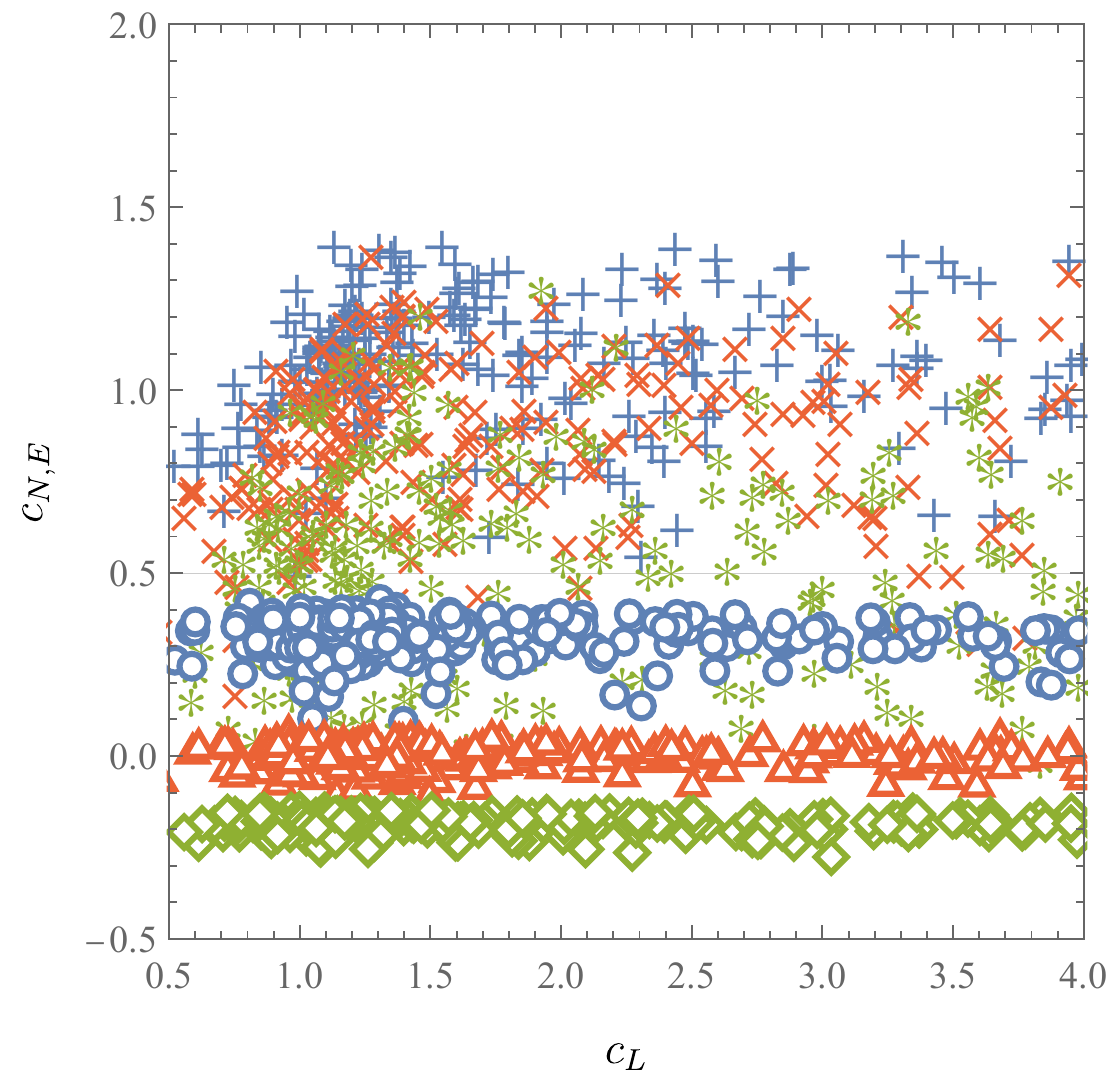}
  \caption{The distribution of 5D fermion mass parameters, $c_i$, for Dirac (left) and Majorana (right) neutrinos. The closed shapes (line symbols) denote $c_{E_i} (c_{N_i})$ values, and the colors blue, red and green correspond to the largest, next-to-largest and smallest $c_{N,E}$ parameters, respectively.}
  \label{fig:c_parameters}
\end{figure*}

The phenomenological predictions of the 5D model are obtained by taking the 5D parameters to be $\sim\mathcal{O}(1)$. Hierarchies in 4D parameters are then generated as a consequence of localisation in the extra dimension. The relevant 5D parameters in the lepton sector are
\begin{equation}
  c_{L_i}, \quad {c_{E_i}}, \quad {c_{N_i}}, \quad y^{(5)}_{e, ij}, \quad y^{(5)}_{\nu, ij}, \quad
  \hat{y}_{N,ij}\,.
\end{equation}
These parameters are then constrained by the nine observables
\begin{equation}
  m_{e_i}, \quad \Delta m^2_{\upnu,ij}, \quad \theta_{ij}, \quad \delta \,,
\end{equation}
where $m_{e_i}$ are the charged lepton masses, $\Delta m^2_{\upnu,ij}$ are the neutrino mass-squared differences, $\theta_{ij}$ are the PMNS mixing angles and $\delta$ is the PMNS Dirac phase.

We perform a numerical scan of the $c_i$ and $y^{(5)}_{e,\nu}$, $\hat{y}_N$ parameter space. The $c_i$ are restricted to satisfy $|c_{L,E,N}|\lesssim 4$, which ensures that the bulk fermion masses remain below the 5D cutoff. For the Majorana case, the $c_{N_i}$ are also restricted to be in the range $[0, 1.4]$ to ensure that the effective Majorana masses remain below the IR scale but sufficiently greater than the Dirac masses (at least $1$ eV). The 5D Yukawa couplings are restricted to satisfy $0.01\leq |y_{e,\nu}^{(5)}|,\hat{y}_N\leq 3$. To increase the efficiency, the scan is done in two stages. First, $y_{e,\nu}^{(5)}$ are fixed to random values, and the $c_i$ and $\hat{y}_N$ fit to the experimental values of the lepton masses by performing a $\chi^2$ minimization using Minuit~\cite{James:1975dr}.  In the second stage, all parameters are floated to fit both the masses and PMNS matrix elements using the values from the first stage as initial seeds for the minimization routine. Finally, we discard points for which $\chi^{2}/n_\mathrm{d.o.f} > 1$, where $n_\mathrm{d.o.f}=9$.

The charged lepton $\overline{\mathrm{MS}}$ masses are run up to the PQ-breaking scale, $10^{10}$ GeV, in order to compare with the model predictions.\footnote{To improve the numerical stability of the fit, we use an enlarged uncertainty of 0.1\% for the charged lepton masses instead of the experimental uncertainty.} The PMNS matrix elements do not run significantly and we use the low-scale values. The neutrino mass differences and PMNS angles and phase are taken from the fit in Ref.~\cite{2007.14792}, assuming a normal hierarchy for the neutrinos. Qualitatively similar results are obtained for an inverted mass hierarchy spectrum.

In the following two sections we present the results for the Dirac and Majorana neutrino cases. As a benchmark point, we assume the following parameter values: $\sigma_0 = 0.1, \lambda = 0.1, \Delta = 10$, $\tan\beta=3$, $k z_{\rm IR} = 10^{7}$, and $k = M_{P}$, which leads to an axion decay constant $F_a \simeq 8.12\times 10^9$ GeV (corresponding to an axion mass, $m_a\simeq 7\times 10^{-4}$ eV~\cite{GrillidiCortona:2015jxo}). The value of $\Delta$ is chosen to sufficiently suppress the axion profile on the UV brane and therefore solve the axion quality problem~\cite{Cox:2019rro}.


\subsection{Dirac neutrinos}

In the Dirac neutrino case, $\hat{y}_N=0$. Given our assumption of $\mathcal{O}(1)$ 5D Yukawa couplings, the neutrino mass scale is solely determined by the $c_L$ and $c_N$ parameters. This case is similar to the quark sector studied in \cite{Bonnefoy:2020llz}. 

Figure~\ref{fig:c_parameters} (left) shows the range of $c_i$ parameters that can produce the measured values of the lepton masses and PMNS angles/phase. Note that for the right-handed charged leptons $|c_{E_i}| \lesssim 0.5$, while for the right-handed neutrinos $c_{N_i} \gtrsim 1$. This simply reflects the fact that the neutrino masses are much smaller than the charged lepton masses. Furthermore, the structures seen in figure~\ref{fig:c_parameters} for both the charged leptons and neutrinos can be understood from the fact that the effective Dirac mass term depends only on $\min(c_L,c_{N/E})$ (see \cref{eq:Dirac-mass-approx}). The hierarchies in the charged leptons can also clearly be seen, whereas the neutrinos need not be hierarchical. 

Since the Higgs VEV profile is flat, the overlap between the left and right-handed fermions is responsible for generating the hierarchies in the 4D effective Yukawa couplings. The $c$ parameter ranges in figure~\ref{fig:c_parameters} correspond to the left (right)-handed fermions being localized towards the UV (IR) branes. Finally, for $c_L<0$ the overlap integral in \eqref{eq:Dirac-mass} is suppressed relative to the electroweak VEV by a factor of $(kz_{IR})^{-n}$, with $n>1/2$; hence, there are no solutions in this region as the charged lepton masses would be too small.

The axial-vector axion--neutrino couplings are shown in Figure~\ref{fig:couplings} (top panel), for $m_{\upnu_1}>10^{-6}$\,eV. The flavor-diagonal couplings are approximately $c_\upnu^A\simeq 2\times 10^{-5}$. Only the coupling $(c_\upnu^A)_{33}$ is shown in the figure, but $(c_\upnu^A)_{11}$ and $(c_\upnu^A)_{22}$ are of the same order of magnitude. In addition, there are flavor non-diagonal $c_\upnu^A$ couplings which are much smaller ($\lesssim 10^{-6}$). Almost identical values are obtained for the off-diagonal vector couplings (note that the diagonal vector couplings are unphysical, up to electroweak anomalies). 

These axion-neutrino couplings can, in principle, be constrained by astrophysical neutrinos scattering off relic axions~\cite{Kelly:2018tyg}, particle emission in double-$\beta$ decay experiments~\cite{KamLAND-Zen:2012uen}, or Planck satellite measurements~\cite{Friedland:2007vv}. To compare with the experimental bounds, we first convert the axial-vector couplings to axial couplings, $g_{a\nu\nu} \sim c_\upnu^A m_\nu/F_a$ with ${\cal L}\supset g_{a\nu \nu} a\bar\nu\gamma_5\nu$. However, the $m_\nu/F_a$ factor suppresses the predicted $g_{a\nu\nu}$ couplings to be well below the current most stringent experimental limit $g_{a\nu\nu}\lesssim 10^{-7}$~\cite{Kelly:2018tyg}.


\subsection{Majorana neutrinos}

In the Majorana neutrino case, we consider the parameter space where $\|M_M^{-1} m_D^T\|\ll1$ and the neutrino mass hierarchy is partially generated by the seesaw mechanism. Nevertheless, the $c_N$ parameters have an important role, since they determine the scale of $M_M$ and $m_D$ via \cref{eq:Majorana-mass,eq:Dirac-mass}. We utilise the mixing matrix in \cref{eq:mixing-matrix}, which gives an excellent approximation in the parameter space we consider. 

The results of the scan are shown in Figure~\ref{fig:c_parameters} (right). Notice that the distribution of the $c_E$ parameters is similar to the Dirac neutrino case. On the other hand, the $c_N$ values are smaller compared to the Dirac case, since part of the neutrino hierarchy is now obtained from the seesaw mechanism. The range of $c$ values in Figure~\ref{fig:c_parameters} again corresponds to left (right)-handed fermions localized on the UV (IR) brane, and to composite right-handed fermions in the dual 4D theory.

\begin{figure}[t]
\centering
\includegraphics[width=0.47\textwidth]{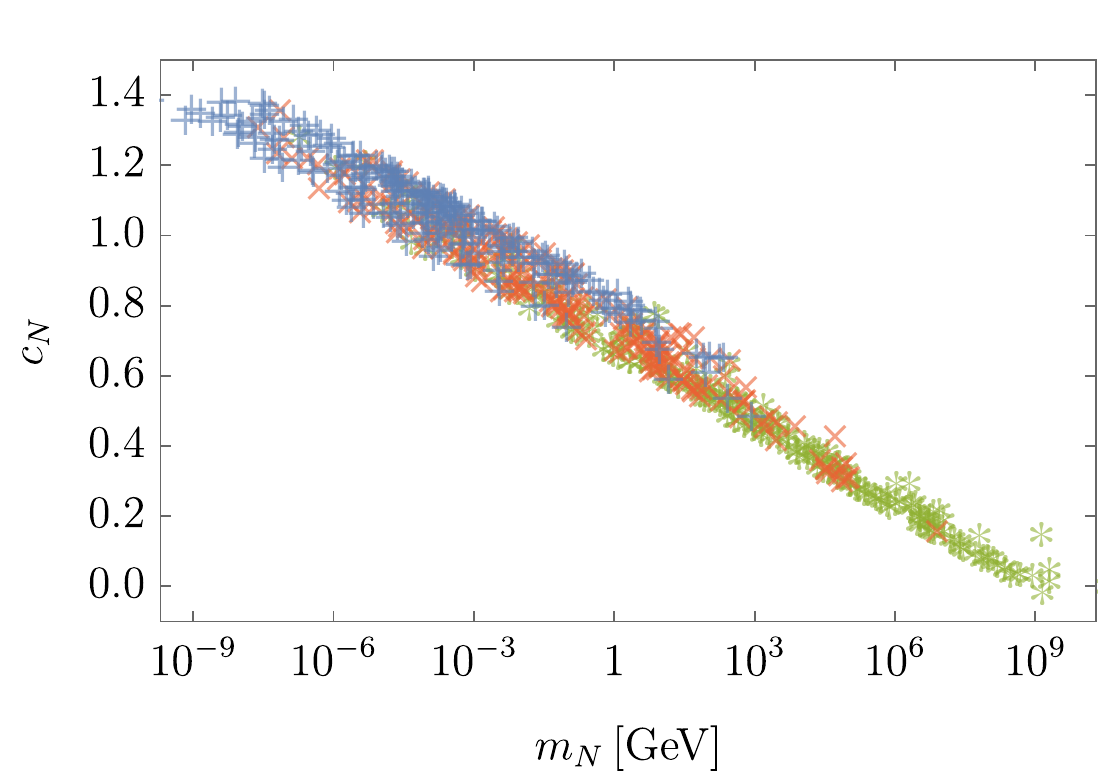}
\caption{The range of sterile neutrino masses arising from the 5D mass parameters $c_{N_i}$. The largest, next-to-largest and smallest $c_N$ parameters are represented by $+$ (blue), $\times$ (red) and $\ast$ (green), respectively. Only sterile neutrino masses below the benchmark IR scale = $2.4 \times 10^{11}$ GeV are shown.}
\label{fig:Sterile_masses}
\end{figure}

Interestingly, and as discussed below \cref{eq:Majorana-mass-approx}, sterile neutrino masses that are hierarchically smaller than the IR scale are naturally obtained for $c_N>0$. This can be clearly seen in Figure~\ref{fig:Sterile_masses}. The assumption that the 5D parameters are $\mathcal{O}(1)$ with flat priors results in a preference for light sterile neutrinos.

\begin{figure}[t]
  \centering
  \begin{subfigure}{.48\textwidth}
    \centering
    \includegraphics[width=\textwidth]{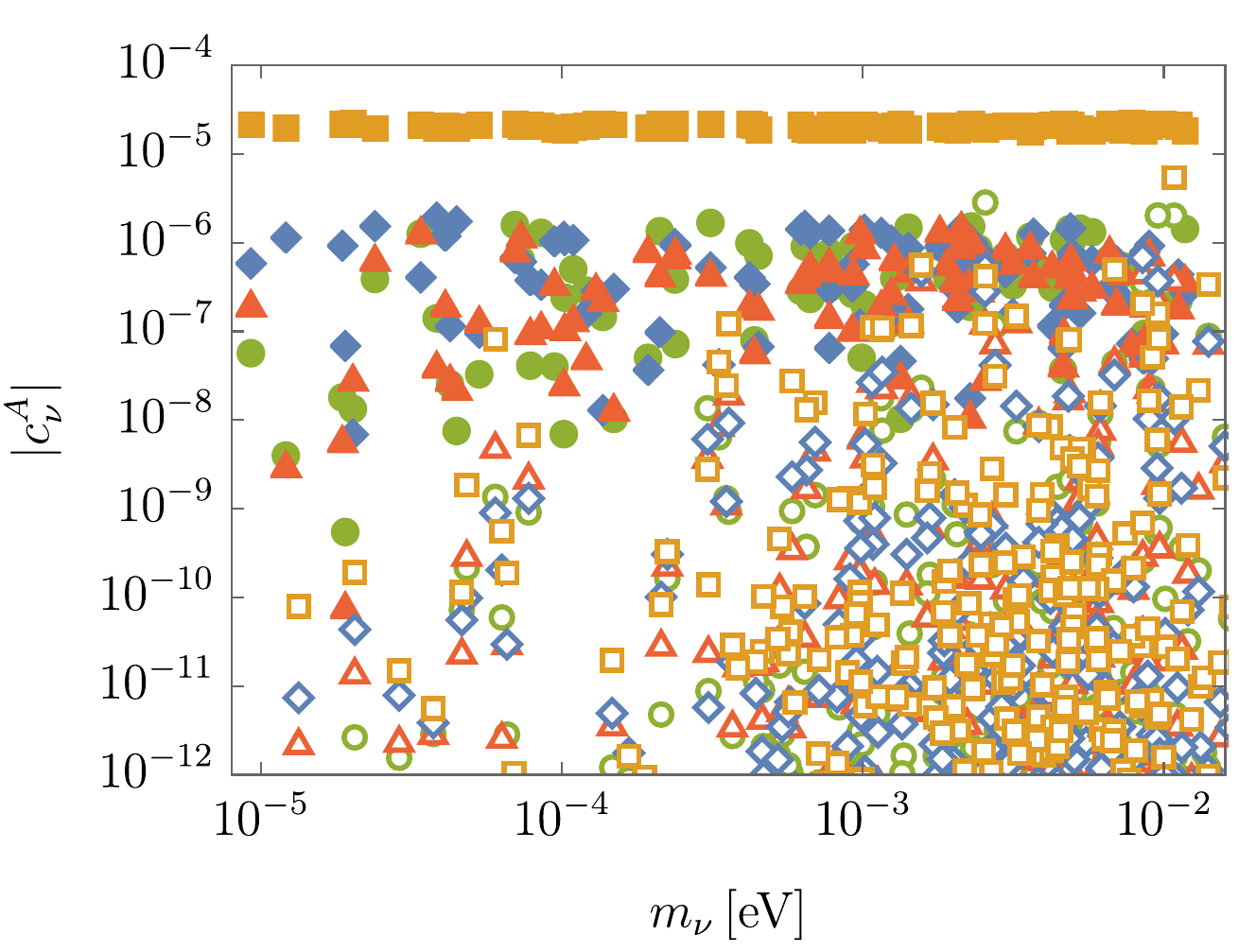}
  \end{subfigure}
  \begin{subfigure}{.48\textwidth}
    \centering
    \includegraphics[width=\textwidth]{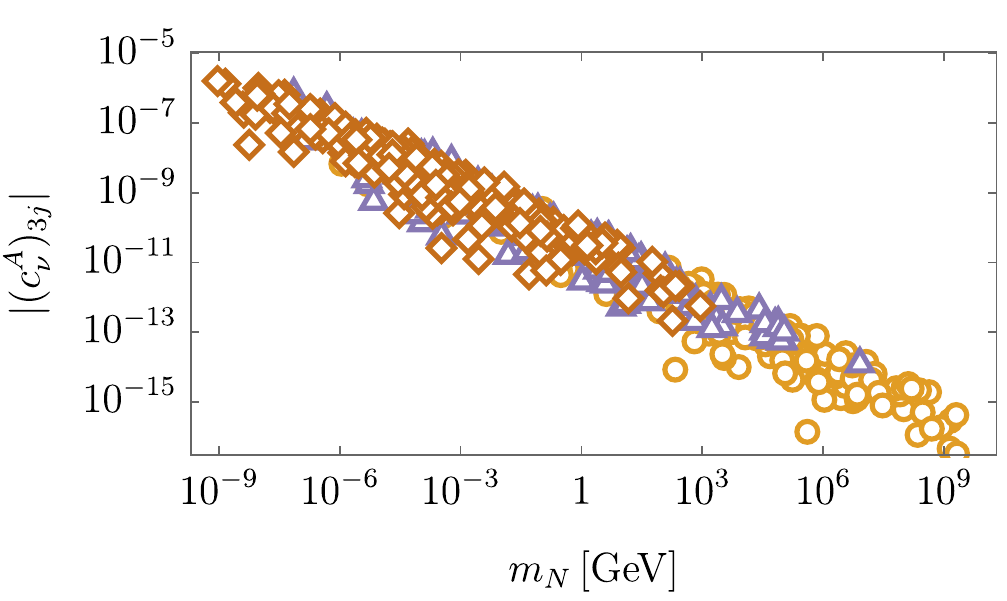}
  \end{subfigure}
  \caption{The axion--neutrino couplings for both the Dirac and Majorana neutrino cases. The top figure shows the active--active axial-vector coupling, $c_\upnu^{A}$, as a function of the lightest active neutrino mass. The solid (open) shapes denote the Dirac (Majorana) couplings. The diagonal coupling is $(c_\upnu^A)_{33}$ (orange square), while the off-diagonal couplings are $(c_\upnu^A)_{12}$ (green circle), $(c_\upnu^A)_{13}$ (red triangle), and $(c_\upnu^A)_{23}$ (blue diamond). The bottom figure shows the active--sterile axial-vector coupling $(c_\upnu^A)_{3j}$  $(j=4,5,6)$ in the Majorana case as a function of the sterile neutrino mass. The lightest, next-to-lightest and heaviest sterile neutrinos are denoted by diamonds (brown), triangles (purple) and circles (orange), respectively.}
  \label{fig:couplings}
\end{figure}

Finally, we discuss the axion--neutrino couplings. The axial-vector couplings are shown in Figure~\ref{fig:couplings}. The vector couplings are similar in magnitude and not shown. The flavor-diagonal axial-vector couplings are much smaller than in the Dirac neutrino case (the diagonal vector couplings are identically zero). This is because, neglecting active-sterile mixing, they depend only on the left-handed profiles, $f^0_{L_{iL}}$, which are UV localized and have a small overlap with the IR-localized $f_{a_u}$. The active--sterile axial-vector couplings are generated through the active--sterile mixing and hence are suppressed for large sterile masses, as seen in figure~\ref{fig:couplings} (bottom). A similar range is found for $(c_\upnu^A)_{1j}$ and $(c_\upnu^A)_{2j}$ (not shown). These axial-vector couplings can be converted to axial couplings $g_{a\nu N} \sim c_\upnu^A m_N/F_a$, with ${\cal L}\supset g_{a\nu N} a\bar\nu\gamma_5N$. Again, the suppression factor $m_N/F_a$ means the predicted couplings are well below current experimental limits~\cite{Kelly:2018tyg}.

\section{Conclusion}
\label{sec:conclusion}

We have presented a 5D model that simultaneously addresses several problems associated with the Standard Model flavor structure, neutrinos and axion physics. The VEV of a 5D complex scalar field spontaneously breaks the Peccei-Quinn symmetry, giving rise to the axion as a Nambu-Goldstone boson with an axion decay constant determined by the IR scale. This setup automatically addresses the axion quality problem by suppressing the axion profile near the UV brane, where there can be explicit (non-QCD) violations of the PQ symmetry.

The 5D model also explains the Standard Model fermion mass hierarchy and flavor structure, while allowing for either Dirac or Majorana neutrinos. This is done by introducing right-handed neutrinos charged under the PQ symmetry. In the Majorana neutrino case, the origin of the Majorana mass scale is associated with explicit, Planck-scale PQ symmetry violation on the UV brane. Nevertheless, by localizing the right-handed neutrino profiles towards the IR brane, hierarchically small sterile neutrino masses can be generated, offering a mechanism to naturally extend the applicability of the seesaw mechanism to much lower mass scales. These light sterile neutrino states may be observable in astrophysical or laboratory experiments~(see e.g. \cite{Bolton:2019pcu}).

In the Dirac case, tiny effective 4D neutrino Yukawa couplings arise from the exponentially small overlap between the left and right-handed neutrino profiles. The axion and neutrino profile structure leads to specific predictions for the axion--neutrino couplings; however, these are well below current experimental limits and would require a substantial improvement in experimental sensitivity to be probed.

The holographic dual 4D description suggests some new strong dynamics with accidental PQ and lepton number symmetries that confines at an intermediate scale $\gtrsim 10^{10}$ GeV and gives rise to a composite axion and composite sterile neutrinos. The light sterile neutrinos result from the suppressed transmission of the explicit lepton number breaking to the composite sector, similar to the setup considered in \cite{Gherghetta:2003hf}. It would be interesting to construct the underlying 4D theory (along the lines studied in \cite{Gherghetta:2020ofz}). Nevertheless, the 5D model provides a complete framework that connects axion and neutrino physics to the Standard Model flavor structure and further motivates ongoing experimental searches for axions and sterile neutrinos.


\begin{acknowledgments}
\subsection{Acknowledgments}
The work of P.C. is supported by the Australian Government through the Australian Research Council. The work of T.G. and M.N. is supported in part by the Department of Energy under Grant DE-SC0011842 at the University of Minnesota, and T.G. is also supported by the Simons Foundation. T.G. acknowledges the Aspen Center for Physics which is supported by the National Science Foundation grant PHY-1607611, where part of this work was done. 
\end{acknowledgments}


\bibliography{citations}

\end{document}